\documentclass[aps,prl,twocolumn,superscriptaddress,footinbib]{revtex4-2}
\pdfoutput=1
\usepackage{amsmath}
\usepackage{graphicx}
\usepackage{bbm}
\usepackage{amssymb}
\usepackage{enumitem}
\usepackage{mathrsfs}
\usepackage[linktocpage=true]{hyperref}

\newcommand{\ex}{\mathrm{e}}
\newcommand{\diff}{\mathrm{d}}
\newcommand{\R}{\mathbb{R}}
\newcommand{\vol}{\mathrm{vol}}

\newcommand{\varphinew}{\phi}

\usepackage{color}

\newcommand{\ii}{\mathrm{i}}

\newcommand{\Z}{\mathbb{Z}}
\newcommand{\rcharge}{q}

\begin{document}

\title{Equivariant localization in supergravity}

\author{Pietro Benetti Genolini}
\affiliation{Department of Mathematics,
King's College London, Strand, London, WC2R 2LS, U.K.}
\author{Jerome P. Gauntlett}
\affiliation{Blackett Laboratory, Imperial College, Prince Consort Road, London, SW7 2AZ, U.K.}
\author{James Sparks}
\affiliation{Mathematical Institute, University of Oxford, Woodstock Road, Oxford, OX2 6GG, U.K.}

\begin{abstract}
\noindent  We show that supersymmetric supergravity solutions 
with an R-symmetry Killing vector are equipped with a set of equivariantly closed 
forms. Various physical observables may be expressed as
integrals of these forms, and then evaluated using the Berline-Vergne-Atiyah-Bott
fixed point theorem. We illustrate with a variety of holographic examples, including 
on-shell actions, black hole entropies, central charges, and scaling dimensions of operators.  
The resulting expressions depend only on topological 
data and the R-symmetry vector, and hence may be evaluated without solving
the supergravity equations. 
\end{abstract}

\maketitle

\section{Introduction}\label{sec:intro}
Starting with the seminal work of \cite{Duistermaat:1982vw}, \cite{Witten:1982im}, 
localization has proved to be an invaluable tool both in mathematics and in
supersymmetric quantum field theory. These ideas have notably been 
applied to the infinite-dimensional supersymmetric path integral (see \cite{Pestun:2016zxk} for a review, and \footnote{Attempts have also been 
made to apply localization to the supergravity path integral, see {\it e.g.}
\cite{Dabholkar:2011ec, Dabholkar:2014wpa}. }), but localization in finite dimensions also plays a role. 
For instance, the localization over instanton moduli spaces in  \cite{Nekrasov:2002qd}, or in the analysis of anomalies and anomaly polynomials. 
In this letter we show that 
localization also 
plays a role in  supergravity. The general structure we uncover explains,
unifies and generalizes many previous results in the literature. 

The supergravity theory or relevant part of the geometry will be
assumed to have even dimension $d=2n$. 
We require that supersymmetric solutions 
are equipped with an R-symmetry Killing vector $\xi$, constructed 
as a bilinear in the Killing spinor $\epsilon$
\begin{align}\label{xidef}
\xi\equiv \bar{\epsilon}\gamma^\mu\gamma_*\epsilon\, \partial_\mu\, .
\end{align}
Here $\gamma_*$ is either $\mathbbm{1}$ or the chirality operator, depending 
on the theory considered, with $\gamma_\mu$ generating the Clifford algebra, 
so $\{\gamma_\mu,\gamma_\nu\}=2g_{\mu\nu}$. 
With an R-symmetry, the space of solutions to the Killing spinor equation generically has
one complex dimension,
generated by $\epsilon$, and one can then argue 
(see {\it e.g.} \cite{Ferrero:2021etw}) that $\mathcal{L}_\xi \epsilon = \ii \rcharge \epsilon$,
 where we refer to the constant $\rcharge$ as the {R-charge}. 

In such a set-up it is natural to introduce the equivariant 
exterior derivative
\begin{align}\label{dxi}
\diff_\xi \equiv \diff - \xi\lrcorner\, .
\end{align}
This acts on differential forms and squares to minus the Lie derivative 
$\diff_\xi^2 = -\mathcal{L}_\xi$. Indeed, in this setting differential forms of 
degree $r$
are constructed naturally as bilinears 
 $\Psi_r\equiv \bar\epsilon \gamma_{(r)}\gamma_*\epsilon$, where 
$\gamma_{(r)}\equiv \frac{1}{r!}\gamma_{\mu_1\cdots\mu_r}
\diff x^{\mu_1}\wedge\cdots\wedge\diff x^{\mu_r}$.  
From our above assumptions on $\xi$ we immediately have 
$\mathcal{L}_\xi \Psi_r=0$.  
The framework of $G$-structures 
and intrinsic torsion implies 
that the Killing spinor equation for $\epsilon$ 
may be recast as a set of 
differential and algebraic equations on 
bilinear forms and  fields in the theory  (see \cite{Gauntlett:2002sc}). 
In general these equations involve bilinears constructed 
using the charge conjugate $\bar\epsilon^c$ as 
well as the Dirac conjugate $\bar\epsilon$, but 
only the latter will appear in the present work; 
the former are charged under $\xi$ when $\rcharge\neq 0$, and 
are thus not invariant under $\mathcal{L}_\xi$. 

In the sequel we will be interested in finding 
polyforms $\Phi$, constructed as polynomials in the bilinears 
$\Psi_r$ and  fields in the supergravity theory,
which by virtue of the Killing spinor equation and equations of motion satisfy
the equivariantly closed condition 
\begin{align}\label{eclosed}
\diff_\xi \Phi = 0\, .
\end{align}
Notice that since $\xi$ is itself a bilinear, the bilinear 
degree of the $(r-2)$-form component $\Phi_{r-2}$ is necessarily one more than $\Phi_{r}$. 
Moreover, since only the Dirac bilinears $\Psi_r$ are used to 
construct $\Phi$, we will only need to impose a  subset 
of the supersymmetry equations and equations of motion to ensure \eqref{eclosed} holds. 
In this sense such structures exist  ``partially off-shell,'' 
and this will also play a role. 
As we will see, a number of such polyforms can exist in a given theory.

Given $\Phi$ satisfying \eqref{eclosed} we may 
integrate it over a $\xi$-invariant closed submanifold $M$, and 
apply the  Berline-Vergne-Atiyah-Bott \cite{BV:1982} \cite{Atiyah:1984px}
fixed point formula.  Denoting an embedded connected 
component of the fixed point set as $f:F\hookrightarrow M$, where $\xi=0$, we have
\begin{align}\label{Philocalize}
\int_M \Phi = \sum_{\substack{F \\ \mathrm{codim} F = 2k}} \frac{1}{d_F}\frac{(2\pi)^k}{\prod_{i=1}^k\epsilon_i}\int_F \frac{f^*\Phi}{\prod_{i=1}^k\left[1+\frac{2\pi}{\epsilon_i}c_1(L_i)\right]}\, .
\end{align}
Here for simplicity\footnote{This assumption covers all of the examples we consider here. When the normal bundle does not split into 
a sum of line bundles, there is a more general formula which can be used to carry out analogous computations.} 
we have assumed that the weights $\epsilon_i$ 
of the linear action of $\xi$ on the normal bundle $N_F$ to $F$ in $M$ are generic, so that 
$N_F=\bigoplus_{i=1}^k L_i$ splits as a sum of complex line bundles, 
with $c_1(L_i)$ denoting the first Chern classes. The normal 
space to a generic point on $F$ is taken to be $\R^{2k}/\Gamma$, 
where $\Gamma$ is a finite group of order $d_F\in\mathbb{N}$, so that 
\eqref{Philocalize} also applies to orbifolds. 

What is perhaps surprising is that many BPS physical quantities 
take the form \eqref{Philocalize}, and furthermore this structure then allows one to 
evaluate ``off-shell,'' without imposing or solving 
the supergravity equations. In the remainder of this letter we give various illustrative
examples, both recovering known results and giving some new results,
leaving comments on further applications and generalizations 
for the discussion section. 

\section{4d minimal gauged supergravity}\label{sec:2}

We consider supersymmetric solutions 
to $4d$, $\mathcal{N}=2$ minimal gauged supergravity. 
The bosonic content is Einstein-Maxwell theory with a negative cosmological constant. In Euclidean signature 
the holographically renormalized action is
\begin{align}\label{4dI}
I =  - \frac{1}{16\pi G_4}\bigg\{&\int_M (R_g+6-F^2)\, \vol_g +\int_{\partial M} 
2\mathcal{K} \, \vol_h \nonumber\\
&  -\int_{\partial M}(4+R_h)\, \vol_h \bigg\}\, .
\end{align}
Here $M$ is a four-manifold, with
boundary $\partial M$ with induced metric $h$, $\vol$ denote Riemannian volume 
forms, $R$ Ricci scalars, and $F=\diff A$ is the Maxwell field strength. The 
Gibbons-Hawking-York  term involves the trace of the extrinsic curvature $\mathcal{K}$,
 while the final counterterm renormalizes the on-shell action 
for asymptotically locally $AdS$ solutions \cite{Emparan:1999pm}. The Newton 
constant in dimension $d$ will be denoted $G_d$. 

Supersymmetric solutions to the Euclidean theory were analyzed in 
\cite{BenettiGenolini:2019jdz}. From the Killing spinor $\epsilon$ one 
may construct the following real bilinear forms
\begin{align}\label{4dbilinears}
& S \equiv  \bar{\epsilon}\epsilon\, , \   P   \equiv \bar{\epsilon}\gamma_5 \epsilon\, ,  \ 
  \xi^\flat  \equiv -\ii \bar{\epsilon}\gamma_{(1)}\gamma_5\epsilon\, , 
\   U   \equiv \ii \bar{\epsilon}\gamma_{(2)}\epsilon\, .
\end{align}
Here $\gamma_5\equiv \gamma_{1234}$, and the one-form $\xi^\flat$ 
is dual to $\xi$, given by \eqref{xidef} with $\gamma_*=-\ii \gamma_5$. 
We then define the polyform
\begin{align}
\Phi =  & \ \Phi_4+\Phi_2+\Phi_0 \nonumber\\ 
\equiv &  \ 
(3\vol_g+F\wedge *F) + (  U + SF -  P *F) - SP\, ,
\end{align}
where $*$ denotes the Hodge dual. 
As indicated in the introduction, this is a polynomial in 
the bilinear forms \eqref{4dbilinears} and 
supergravity form fields, with $\Phi_{2j}$
being  degree $2-j$ in bilinears. 
The Killing spinor equation implies 
certain differential conditions on these forms, 
and using the results in \cite{BenettiGenolini:2019jdz} 
one easily shows that $\Phi$ satisfies the equivariantly 
closed condition \eqref{eclosed}. 

Using the Einstein equation the on-shell 
action \eqref{4dI} is given by
\begin{align}
I = \left[\frac{1}{(2\pi)^2}\int_M \Phi\right]\frac{\pi}{2G_4} + 
\mbox{boundary terms}\, .
\end{align}
We cannot directly apply \eqref{Philocalize} in this case, 
since $M$ has a boundary. 
However, following the way in which the fixed point theorem is proven, we notice that the integrand $\Phi$
is exact on the complement of the fixed point set and the boundary. Then, 
assuming\footnote{As discussed in \cite{BenettiGenolini:2019jdz}, we are then considering supergravity solutions associated 
with the general field theory results of \cite{Closset:2012ru}. It would be interesting to know if there are  
solutions when this assumption is relaxed.}  $\xi$ has  no fixed points on the boundary one can integrate by 
parts, leading to a boundary term on $\partial M$ together 
with contributions around the fixed point set in the interior of $M$. 
The (divergent) boundary contribution exactly cancels with the boundary integrals in \eqref{4dI}.
This was shown by explicit 
computation in \cite{BenettiGenolini:2019jdz}, but is also expected since the result should be Weyl invariant
and for this theory there is no such boundary quantity; equivalently we cannot construct a finite Weyl invariant counterterm \footnote{We thank K. Skenderis for this comment.}.
Thus, from \eqref{Philocalize} the remaining fixed point contribution in the interior of $M$ gives
\begin{align}\label{I4dexpand}
I = \bigg\{\sum_{\substack{\mathrm{fixed}\\ \mathrm{points}}}\frac{ \Phi_0}{\epsilon_1\epsilon_2} +
\sum_{\substack{\mathrm{fixed}\\ \Sigma}} \int_\Sigma 
\frac{\Phi_2}{2\pi \epsilon_1}-\frac{\Phi_0 c_1(L)}{\epsilon_1^2} \bigg\}\frac{\pi}{2G_4}\, .
\end{align}
Here $L$ is the normal bundle to the surface $\Sigma$ in $M$, and we note 
that $\Phi_0$ is necessarily constant over $\Sigma$.  

Following \cite{BenettiGenolini:2019jdz}
we may write $P=S\cos\theta$, where  $\|\xi\|=
S |\sin\theta|$. Since the spinor square norm $S$ is necessarily 
nowhere zero (see \cite{Ferrero:2021etw} for a general argument), 
it follows that at a fixed point set $\theta=0$ or $\pi$, and hence
correspondingly $P=\pm S$,  so that $\epsilon$ is
necessarily of fixed chirality with $\gamma_5\epsilon = \pm \epsilon$ at such a fixed locus.

Examining first an isolated fixed point, on the tangent space we may write $\xi = \sum_{i=1}^2 \epsilon_i\, \partial_{\varphi_i}$, 
where $\partial_{\varphi_i}$ rotate each copy of $\R^2_i$ in $\R^4=\R^2_1\oplus\R^2_2$.  
Notice here that the overall orientation on $\R^4$ is fixed, but the orientations 
of each $\R^2_i$ factor are not; this means that the pair $(\epsilon_1,\epsilon_2)$ 
is only defined up to overall sign.
A local analysis of the bilinears near such a fixed point relates 
$S$ to the norm of the self-dual and anti-self-dual parts of the two-form
$\diff\xi^\flat$, leading to the general formula 
$S=|\epsilon_1\mp \epsilon_2|/2$
\cite{BenettiGenolini:2019jdz}. A similar argument leads to 
 $S=-\epsilon_1/2$ for a fixed surface, while 
 formulae in \cite{BenettiGenolini:2019jdz} immediately give 
\begin{align}
\int_\Sigma \Phi_2  = 2\int_\Sigma SF = -\epsilon_1 \int_\Sigma F\, .
\end{align}
A global analysis of spinors near to the fixed surface $\Sigma=\Sigma_\pm$, 
which have charge $\tfrac{1}{2}$ under $A$, then implies \cite{BenettiGenolini:2019jdz}
\begin{align}
\frac{1}{2\pi}\int_{\Sigma_\pm} F = -\frac{1}{2}\int_{\Sigma_\pm}c_1(T\Sigma_\pm)\mp c_1(L)\, ,
\end{align}
where $\int_{\Sigma_\pm}c_1(T\Sigma_\pm)=2(1-g_\pm)$ is the Chern number of the tangent bundle of the Riemann surface $\Sigma_\pm$ of genus $g_\pm$,
and substituting into \eqref{I4dexpand} gives
\begin{align}\label{I4dfinal}
I  = & \ \bigg\{\sum_{\substack{\mathrm{fixed}\\ \mathrm{points}_\pm}}\mp \frac{(\epsilon_1\mp \epsilon_2)^2}{4\epsilon_1\epsilon_2} + 
\sum_{\substack{\mathrm{fixed}\\ \Sigma_\pm}} \int_{\Sigma_\pm} \Big[\frac{1}{2}c_1(T\Sigma_\pm)
\nonumber\\ 
&  \qquad \mp\frac{1}{4}c_1(L)\Big] \bigg\}\frac{\pi}{2G_4}\, .
\end{align}
This is the main result of \cite{BenettiGenolini:2019jdz}, derived 
here as a simple application of \eqref{Philocalize}. 

\section{6d Romans F(4) gauged  supergravity}

A similar structure exists in $6d$ Romans F(4) gauged supergravity, 
where for the Euclidean theory we follow
\cite{Alday:2014rxa, Alday:2015jsa}. We work in an 
Abelian truncation, where in addition to the metric the bosonic content of the
theory has a 
scalar field $X$, Maxwell field strength $F=\diff A$, 
all of which are real, and an imaginary two-form potential~$B$.

 From the Killing spinor $\epsilon$ one 
may construct the following bilinear forms
\begin{align}\label{6dbilinears}
S  \equiv  \bar{\epsilon}\epsilon\, , \   P \equiv \bar{\epsilon}\gamma_7 \epsilon\, , \ \xi^\flat \equiv \bar\epsilon \gamma_{(1)}\epsilon\, , \nonumber \\
Y \equiv \ii \bar\epsilon \gamma_{(2)}\epsilon\, , \ 
\tilde{Y} \equiv \ii \bar\epsilon \gamma_{(2)}\gamma_7\epsilon\, ,
\end{align}
where $\gamma_7\equiv \ii \gamma_{123456}$  and the one-form $\xi^\flat$ 
is dual to $\xi$, given by \eqref{xidef} with $\gamma_*=\mathbbm{1}$.  
Using these we then define the polyform $\Phi=\Phi_6+\Phi_4+\Phi_2+\Phi_0$, where
\begin{align}
\Phi_6 \equiv & \ \tfrac{4}{9}\tfrac{2+3X^4}{X^{2}}\, \vol + \tfrac{1}{3}
X^{-2}F\wedge *F+\tfrac{\ii}{3}B\wedge F\wedge F\, , \nonumber\\
\Phi_4\equiv & \ \tfrac{\sqrt{2}}{3}(XP)X^{-2}*F -\tfrac{2\sqrt{2}}{3}X*\tilde{Y}- \tfrac{\sqrt{2}}{3}F \wedge X^{-1}Y \nonumber\\
& +\tfrac{1}{\sqrt{2}}
(XS){F}\wedge{F} +\tfrac{2\sqrt{2}\ii}{3}(XP)B\wedge F\, , \nonumber\\
\Phi_2 \equiv & \ -\tfrac{2}{3}P Y + \tfrac{2\ii}{3}(XP)^2 B + 2(XS)(XP)F\, ,\nonumber\\
\Phi_0 \equiv & \ \sqrt{2}(XS)(XP)^2\, .
\end{align}
Notice that $\Phi_{2j}$ has bilinear degree $3-j$.
The differential equations satisfied by \eqref{6dbilinears} may be found in 
\cite{Alday:2015jsa}, and using these one can show that $\Phi$ 
satisfies the equivariantly closed condition \eqref{eclosed}.

The on-shell action may be written as
\begin{align}
I = \left[\frac{1}{(2\pi)^3}\int_M \Phi \right]\frac{\pi^2}{2G_6}+\mathrm{boundary\,  terms}\, ,
\end{align}
where the full set of boundary counterterms may be found in \cite{Alday:2014rxa}. 
Similar to the case of $4d$ minimal gauged supergravity, for this theory there are again no
finite Weyl invariant counterterms. Thus, assuming that the fixed point set lies within the interior of $M$,
the boundary terms will cancel leaving only a fixed point contribution in the interior. 
It is then straightforward to write down an explicit expression for the on-shell action using localization.
For brevity, we just give the result for the class of solutions in which 
there are only isolated fixed points:
\begin{align}\label{6dIlocalize}
I = \sum_{\substack{\mathrm{fixed}\\ \mathrm{points}}} \pm
\frac{(\epsilon_1+\epsilon_2+ \epsilon_3)^3}{\epsilon_1\epsilon_2\epsilon_3} 
\frac{\pi^2}{4G_6} \,,
\end{align}
where $\R^6=\oplus_{i=1}^3\R^2_i$, and as commented in the previous section
only the overall orientation is fixed. 
Here we have used the fact that at a fixed point set again $|P|=S$, 
together with the  local analysis in \cite{Alday:2015jsa} which shows that 
$(XS)\mid_{\mathrm{fixed}\, \mathrm{point}} \,  = {(\epsilon_1+\epsilon_2+ \epsilon_3)}/{\sqrt{2}}$, 
in appropriate orientation conventions for each $\R^2_i$. Having chosen such conventions, the orientation 
on $\R^6$ then either agrees with the fixed orientation or not, which we write as a $\pm$ sign in \ref{6dIlocalize}.

Remarkably, we have recovered the conjectured result 
for the on-shell action given in \cite{Alday:2014rxa}. This conjecture was known to hold for 
at least three different families of examples, including the non-rotating black hole solutions in
\cite{Alday:2014fsa} (with hyperbolic space horizons), but here we have obtained a general proof, which goes beyond these examples.

As a further illustration, we show that \eqref{6dIlocalize} correctly gives 
the on-shell action and hence entropy of supersymmetric rotating black hole solutions in this theory. 
We analyse the complex branch of supersymmetric black hole solutions 
studied in \cite{Cassani:2019mms}. The spacetime $M$ has the topology 
$\R^2\times S^4$, and the R-symmetry Killing vector is
\begin{align}
\xi = \sum_{i=1}^2 \ii \frac{\omega_i}{\beta}\partial_{\varphi_i} + \frac{2\pi}{\beta}\partial_{\varphi_3}\, .
\end{align}
Here $\omega_i$, $i=1,2$ are (complex) angular velocity chemical potentials, and we have 
embedded $S^4\subset \R^2_1\oplus\R^2_2\oplus\R$, with 
$\partial_{\varphi_i}$ rotating the $\R^2_i$ factors, $i=1,2$, 
while $\varphi_3=2\pi \tau/\beta$, with $\tau$ the Euclidean 
time circle, of period $\beta$. For generic values of parameters 
the fixed point set consists of the north and south 
poles of the $S^4$ horizon. These give an equal contribution, and
\eqref{6dIlocalize} (with an overall plus sign)
gives the on-shell action
\begin{align}\label{IBHRomans}
I = \frac{\pi [2\pi + \ii (\omega_1+\omega_2)]^3}{4\omega_1 \omega_2 G_6}\, .
\end{align}
This agrees with the result in  \cite{Cassani:2019mms}, 
after taking into account the fact that the $AdS$ radius 
in our conventions is $\ell=3/\sqrt{2}$. 
The entropy 
of the black hole may then be computed via a  Legendre transform, extremizing
\begin{align}\label{SBHRomans}
S = -I - \sum_{i=1}^2 \omega_i J_i - \tfrac{\ell}{3}(-2\pi \ii +\omega_1 + \omega_2)Q\, ,
\end{align}
over the chemical potentials $\omega_i$, 
where $J_i$, $Q$ are the angular momenta and electric charge, respectively.  

\section{$AdS_5\times M_6$ Solutions}

In this section we consider supersymmetric 
$AdS_5$ solutions to $11d$ supergravity, as analysed in 
\cite{Gauntlett:2004zh}. The $11d$ metric takes the warped product form
\begin{align}
\diff s^2_{11} = \ex^{2\lambda}(\diff s^2_{AdS_5}+\diff s^2_{M_6})\, ,
\end{align}
where we take $AdS_5$ to have unit radius, and will assume that 
$M_6$ is compact without boundary. The four-form flux $G$ 
and function $\lambda$ are pull-backs from $M_6$.

Denoting $\epsilon=\epsilon^+_{\mathrm{there}}$ in \cite{Gauntlett:2004zh}, 
we have bilinears on $M_6$:
\begin{align}
1 = \bar\epsilon\epsilon\, , \ \sin\zeta\equiv -\ii\bar\epsilon\gamma_7\epsilon\, , \ 
\xi^\flat \equiv \tfrac{1}{3}\bar\epsilon \gamma_{(1)}\gamma_7\epsilon\, , 
\nonumber \\
Y\equiv -\ii \bar\epsilon \gamma_{(2)}\epsilon\, , \ Y' \equiv \bar\epsilon\gamma_{(2)}\gamma_7\epsilon\, ,
\end{align}
where $\gamma_7\equiv \gamma_{123456}$. There are some 
immediate differences with the applications in the previous sections: 
the spinor necessarily has constant norm, which we take to be 1, 
but there is  instead a warp factor function $\ex^{2\lambda}$. 
We have also normalized the Killing vector $\xi$ so that 
the R-charge is $\rcharge=\tfrac{1}{2}$. In the notation 
of \cite{Gauntlett:2004zh} then $\xi=\partial_\psi$. An important 
role is played by the function
\begin{align}
y \equiv \tfrac{1}{2}\ex^{3\lambda}\sin\zeta\, ,
\end{align}
which was used as a canonical coordinate in \cite{Gauntlett:2004zh}.

We find the following collection of equivariantly closed forms under $\diff_\xi$
\begin{align}\label{Phibils}
\Phi \equiv & \ \ex^{9\lambda}\vol +\tfrac{1}{12}\ex^{9\lambda}*Y-\tfrac{1}{36}
y\, \ex^{6\lambda}Y - \tfrac{1}{162}y^3\, ,\nonumber\\
 \Phi^G \equiv & \ G -\tfrac{1}{3}\ex^{3\lambda}Y'+\tfrac{1}{9}y\, ,\nonumber\\
\Phi^Y \equiv & \ \ex^{6\lambda} Y +\tfrac{1}{3}y^2\, .
\end{align}
We emphasize that closure under $\diff_\xi$ uses the differential 
conditions on the Dirac bilinears only, which is a strict subset of 
the equations in \cite{Gauntlett:2004zh}. Also, since 
$\epsilon$ here has unit norm the bilinear degrees in \eqref{Phibils}
are less transparent, although one could choose to keep this norm arbitrary.  
The $a$ central charge for such a solution is \cite{Gauntlett:2006ai}
\begin{align}\label{aloc}
a = \frac{1}{2(2\pi)^6\ell_p^9} \int_{M_6}\Phi\, ,
\end{align}
where $\ell_p$ is the $11d$ Planck length; this  then
localizes using \eqref{Philocalize}. 

As an example, let us consider the near-horizon 
limit of $N$ M5-branes wrapped on a spindle. 
The full supergravity solutions were constructed in 
\cite{Ferrero:2021wvk}, with $M_6$ being the total 
space of an $S^4$ bundle fibred over a spindle $\Sigma=\mathbb{WCP}^1_{[n_+,n_-]}$.  
The latter is topologically a two-sphere, but with conical deficit angles 
$2\pi(1-1/n_\pm)$ at the poles. 
We write the R-symmetry vector as 
\begin{align}
\xi = \sum_{i=1}^2 b_i \partial_{\varphi_i}+\varepsilon\, \partial_{\varphi_3} \, ,
\end{align}
where $b_i$ and $\varepsilon$ are constants, arbitrary at this stage, which define the Killing vector. 
Here $\partial_{\varphi_i}$ rotate the two copies of $\R^2_i$ in
$S^4\subset \R^2_1\oplus\R^2_2\oplus\R$, while 
$\partial_{\varphi_3}$ is a lift of the vector field that rotates the spindle, where we  use the construction of such a basis in~\cite{Boido:2022mbe}. 

Consider first fixing one of the poles on $\Sigma$, say the plus pole with orbifold group $\Z_{n_+}$, 
and consider a linearly embedded $S^2_i\subset \R^2_i\oplus \R \subset \R^5$ in the (covering space of the) fibre over it. The homology class of this $S^2_i$ is trivial, 
so it follows that
\begin{align}
0 = \int_{S^2_i} \Phi^Y  = \frac{2\pi}{b_i^\pm}\frac{1}{3}\left[(y^+_N)^2 - (y^+_S)^2\right]\, ,
\end{align}
where the $N$ and $S$ subscripts refer to the poles in the fibre sphere $S^4$, 
and $b_i^\pm$ are the weights of the Killing vector at these poles (i.e. the $\epsilon_i$ in \eqref{Philocalize}), which we shall determine 
below. 
This immediately implies that $|y^\pm_N| = |y^\pm_S|$. 

We next consider flux quantization through the fibres $S^4/\Z_{n_\pm}$ 
over the poles of $\Sigma$. This reads
\begin{align}\label{Npm}
N_\pm = & \ \frac{1}{(2\pi \ell_p)^3} \int_{S^4/\Z_{n_\pm}} \Phi^G \nonumber \\ 
 = & \ 
\frac{1}{(2\pi \ell_p)^3}\frac{1}{n_\pm} \frac{(2\pi)^2}{b_1^\pm b_2^\pm }\frac{1}{9}(y^\pm_N - y^\pm _S)\, .
\end{align}
With $N_\pm>0$, this fixes the signs to be $y^\pm_N=-y^\pm_S>0$. Moreover, 
from the homology relation between these cycles we deduce
\begin{align}
N \equiv n_+ N_+ = n_- N_-\, .
\end{align}
The central charge \eqref{aloc} may then also be computed by localizing
\begin{align}
\int_{M_6}\Phi = & \ -{(2\pi)^3}\bigg[\frac{1}{n_+}\frac{(y^+_N)^3-(y^+_S)^3}{162}\frac{1}{(-\varepsilon/n_+) b_1^+ b_2^+} \nonumber \\ 
& \quad  + \frac{1}{n_-}\frac{(y^-_N)^3-(y^-_S)^3}{162}\frac{1}{(\varepsilon/n_-) b_1^- b_2^-}\bigg]\, . 
\end{align}
Using \eqref{Npm} this remarkably simplifies to
\begin{align}\label{ablock}
a = \frac{9[(b_1^+ b_2^+)^2 - (b_1^- b_2^-)^2]}{16\varepsilon} N^3\, .
\end{align}
This takes a ``gravitational block'' form (see \cite{Hosseini:2019iad}), involving a difference of M5-brane anomaly
polynomials in the numerator, one associated to each $\pm$ pole of $\Sigma$. 

In order to evaluate \eqref{ablock} further we need to first describe the fibration 
structure in more detail. The normal bundle to the M5-brane wrapped on 
$\Sigma$ is $N_\Sigma = \mathcal{O}(-q_1)\oplus \mathcal{O}(-q_2)$, 
where in order for the total space to be Calabi-Yau (giving a topological twist) 
we have 
\begin{align}
q_1 + q_2 = n_+ + n_-\, . 
\end{align}
The weights $b_i^\pm$ may then be computed using the results 
in  \cite{Boido:2022mbe}. We have 
\begin{align}\label{b1b2sum}
b_1^\pm + b_2^\pm = 1 \mp \frac{\varepsilon}{n_\pm}\, ,\quad 
b_i^+ - b_i^- = -\frac{q_i}{n_+n_-}\varepsilon\, ,
\end{align}
the first equation coming from the charge of the holomorphic $(3,0)$-form 
on the Calabi-Yau, and the second equation being (3.24) of \cite{Boido:2022mbe} 
(with $q_i=-p_i^{\mathrm{there}}$). We may then solve these constraints 
by introducing new variables $\varphinew_i$ via
\begin{align}\label{bchangephi}
b_i^\pm = \tfrac{1}{2}\big(\varphinew_i \mp \frac{q_i}{n_+n_-}\varepsilon\big)\, ,
\end{align}
with the constraint 
\begin{align}\label{phicon}
\varphinew_1 + \varphinew_2 = 2 + \frac{n_+-n_-}{n_+n_-}\varepsilon\, .
\end{align}
Our final central charge is then
\begin{align}\label{a6dfinal}
a = -\frac{9[(q_2\varphinew_1 + q_1\varphinew_2)(q_1q_2\varepsilon^2+n_+^2n_-^2\varphinew_1\varphinew_2)]}{64n_+^3n_-^3}N^3\, .
\end{align}
This derives the conjectured gravitational block formula in \cite{Faedo:2021nub}, 
where we have corrected the overall sign. 
In that reference it was shown that extremizing  $a$ over the variables 
$\varphinew_i$ (subject to \eqref{phicon}) gives the central charge 
as well as determines the R-symmetry Killing vector 
of the explicit supergravity solutions constructed in \cite{Ferrero:2021wvk}. 
Moreover, \eqref{a6dfinal} agrees \emph{off-shell} with the 
trial $a$-function in field theory, obtained by integrating 
the M5-brane anomaly polynomial over the spindle. 

This extremization is also explained by our gravity formalism: 
we have imposed a subset of the supersymmetry equations 
to obtain \eqref{a6dfinal}. Substituting this back into the action, 
one then extremizes the resulting expression over any remaining 
degrees of freedom to obtain the on-shell result. 
This is the same general idea used in \cite{Martelli:2006yb,Couzens:2018wnk}, and will be discussed further 
in the present context in \cite{BenettiGenolini:2023yfe,BenettiGenolini:2023ndb}. 

Finally, let us consider the conformal dimensions  of chiral primary operators in the dual SCFT that are
associated with M2-branes 
wrapped over the copies of the spindle $\Sigma_{N}$, $\Sigma_S$ 
at the poles of the $S^4$. We have the general result \cite{Gauntlett:2006ai}
\begin{align}
\Delta(\Sigma) = \frac{1}{(2\pi)^2\ell_p^3}\int_{\Sigma} \ex^{3\lambda} Y'\, ,
\end{align}
where $\Sigma$  is calibrated by $Y'$. Since the latter is always closed 
when restricted to $\Sigma$, $\Phi^G$ in \eqref{Phibils} defines an equivariantly closed form on $\Sigma$, and  localization gives
\begin{align}
& \Delta(\Sigma_N) =   \frac{-1}{(2\pi)^2\ell_p^3}\Big[\frac{1}{n_+}\frac{2\pi}{(-\varepsilon/n_+)}\frac{y^+_N}{3}+\frac{1}{n_-}\frac{2\pi}{(\varepsilon/n_-)}\frac{y^-_N}{3}\Big]\, , \nonumber\\
  & \qquad = \frac{3(b_1^+b_2^+ - b_1^-b_2^-) }{2\varepsilon}N = -\frac{3(q_2\varphinew_1 + q_1\varphinew_2)}{4n_+n_-}N\, ,
\end{align}
with the same result for $\Sigma_S$ (up to orientation). 
Evaluating this on the extremal values $\varphinew_i^*$, $\varepsilon^*$, 
one can verify the result agrees with that computed using the explicit 
supergravity solutions in \cite{Ferrero:2021wvk}. 

Similar calculations reproduce central charges and scaling dimensions 
for many other classes of $AdS_5\times M_6$ solutions, including
all those in \cite{Gauntlett:2004zh}, the M5-branes wrapped 
on general Riemann surfaces in \cite{Bah:2012dg}, and also 
 new results for which explicit supergravity solutions have not
been constructed \cite{BenettiGenolini:2023yfe,BenettiGenolini:2023ndb}. One only needs 
to input topological data for the solutions, as we have done above.

 \section{Discussion}
 The general structure we have uncovered in supergravity is ripe for many further applications and generalizations. 
We certainly expect analogous results to hold for more general supergravity theories,
including coupling to matter multiplets, and including higher derivative corrections 
(see in particular \cite{Bobev:2021oku, Genolini:2021urf, Hristov:2021qsw}). 

We have focused on supergravity geometries in even dimensions, but generalizations to odd dimensions
are also possible. This should lead to a derivation of entropy functions for 
supersymmetric black holes in diverse dimensions, generalizing our derivation of 
 the on-shell action \eqref{IBHRomans} and entropy function \eqref{SBHRomans}
  (see also \cite{Cassani:2021dwa}), and also gravitational block formulae that have been discovered
 in GK geometry \cite{Boido:2022iye,Boido:2022mbe}, and generalizations thereof.   
The latter is very much 
    related to the computation of anomaly polynomials in field theory, and it would be 
    interesting to make 
 contact with the gravitational approach in \cite{Bah:2019rgq}.
 We will report on many of these topics in the near future \cite{BenettiGenolini:2023yfe,BenettiGenolini:2023ndb}.
 
\section*{Acknowledgments}
We thank K. Skenderis for discussions.
This work was supported in part by STFC grants  ST/T000791/1 and 
ST/T000864/1.
JPG is supported as a Visiting Fellow at the Perimeter Institute. 
PBG is supported in part by the Royal Society Grant RSWF/R3/183010.

\bibliographystyle{apsrev}

\bibliography{helical}{}

\end{document}